\documentclass[fleqn,10pt]{wlscirep}
\usepackage{graphicx}
\usepackage{epstopdf}
\title{Multi-resolution community detection in massive networks}

\author[1,*]{Jihui Han}
\author[1,+]{Wei Li}
\author[1,@]{Weibing Deng}
\affil[1]{Complexity Science Center and Institute of Particle Physics, Central China Normal University, Wuhan, 430079, China}

\affil[*]{jh@mails.ccnu.edu.cn}
\affil[+]{liw@mail.ccnu.edu.cn}
\affil[@]{liw@mail.ccnu.edu.cn}

\begin{abstract}
Aiming at improving the efficiency and accuracy of community detection in complex networks, we proposed a new algorithm, which is based on the idea that communities could be detected from subnetworks by comparing the internal and external cohesion of each subnetwork. In our method, similar nodes are firstly gathered into meta-communities, which are then decided to be retained or merged through a multilevel label propagation process, until all of them meet our community criterion. Our algorithm requires neither any priori information of communities nor optimization of any objective function. Experimental results on both synthetic and real-world networks show that, our algorithm performs quite well and runs extremely fast, compared with several other popular algorithms. By tuning a resolution parameter, we can also observe communities at different scales, so this could reveal the hierarchical structure of the network. To further explore the effectiveness of our method, we applied it to the E-Coli transcriptional regulatory network, and found that all the identified modules have strong structural and functional coherence.
\end{abstract}

\begin{document}

\flushbottom
\maketitle

\thispagestyle{empty}

\section*{Introduction}

A wide range of natural and social systems can be described as complex networks \cite{Rubinov20101059,Broder2000309,Palla2007,Onnela01052007,Barabasi2004,Ravasz1551}. Examples include the cell, a network of chemicals linked by chemical reactions, or the Internet, a network of routers and computers connected by physical links.

Most real-world networks are observed to contain communities \cite{Fortunato201075}. Intuitively, a community is a group of nodes which are relatively densely connected to each other within the group but sparsely connected to the nodes in other groups of the network \cite{porter2009communities}. Detecting communities can not only uncover the relations between internal structures and functional behaviours of networks, but also have many practical applications in the domains such as biology, sociology, economics and computer science \cite{10.1371/journal.pone.0039475,10.1371/journal.pone.0014248,Tang20121}. Therefore, it is not surprising that community detection has been so extensively investigated during the past few years \cite{Fortunato201075}. 

In the last decade, lots of methods have been developed to detect the community structure, such as modularity optimization \cite{PhysRevE.69.026113,Newman06062006,PhysRevE.70.066111,PhysRevE.72.027104,FastUnfolding2008}, dynamic label propagation \cite{PhysRevE.76.036106,1367-2630-12-10-103018,JieruiXie-LabelRank-NSW:2013,Lin2014386,Liu2016}, statistical inference \cite{PhysRevE.83.016107,Aldecoa2013surprise}, Potts model\cite{PhysRevE.80.016109,PhysRevE.81.046114}, spectral clustering \cite{PhysRevE.74.036104,Singh2015}, information-theoretic \cite{Rosvall01052007,Rosvall29012008} and topology based \cite{Liu2014,Zalik2015,Chen2016,PhysRevE.72.046108} methods. Among them, one popular group of approaches are based on the optimization of modularity, which could be more or less subject to the resolution problem \cite{Fortunato02012007}. Especially fast ones are label propagation algorithms, in which each node adopts the majority label among its neighbours, and the labels propagate iteratively until there is no change in the network. Due to the frequent tie-breaks and random processing order of nodes, label propagation algorithms usually deliver multiple partitions starting from the same initial condition with different random seeds.

Among all the community detection methods developed so far, it is yet unlikely to obtain a widely accepted definition of community. Hence, we first introduce a quantitative criterion to determine what kind of subnetworks are communities, and then propose a fast algorithm to detect them. The proposed method mainly consists of two steps: initialization and multilevel label propagation. The initialization step is implemented to form some meta-communities by collecting similar nodes. While the multilevel label propagation step, which is similar to the Louvain method \cite{FastUnfolding2008}, consists of two sub-steps: network collapse and label propagation. Firstly, it aggregates nodes that belong to the same meta-community and builds a new network whose nodes represent the meta-communities detected in the previous step. Secondly, it retains or merges meta-communities by comparing their internal and external connections (or weights) through a weighted version of label propagation. These two sub-steps repeat iteratively until all the meta-communities meet our community criterion. As discussed above, our method requires neither optimization of objective function nor any prior information of communities. We tested our algorithm on both synthetic and real-world networks, and compared it with several other popular algorithms. Results show that our algorithm could detect meaningful communities in large networks efficiently and accurately, and it could also uncover the hierarchical structure of the network by tuning a resolution parameter.

\section*{Results}

We evaluated the performance of our method on both synthetic and real-world networks. For synthetic networks, we tested the classical benchmark proposed by Girvan and Newman (GN) \cite{Girvan11062002}, and the well-known benchmark with planted community structure and heterogeneous distributions of node degree and community size proposed by Lancichinetti, Fortunato and Radicchi (LFR) \cite{PhysRevE.78.046110}. As real-world networks have some different topological properties that distinguish them from the synthetic ones, we also tested our method on different kinds of real-world networks, such as social networks and biological networks. To assess its performance, we compared our algorithm with other six popular algorithms listed in Table \ref{algorithms}, in terms of normalized mutual information (NMI) and modularity.

\subsection*{Synthetic Networks}

In this section, we tested our method against synthetic benchmarks. In our algorithm, we set $\lambda = 1$ by default ($\lambda$ can be regarded as a resolution parameter that controls the scale on which we would like to observe the communities in a network, see \textbf{Methods} for detailed discussions). In all tests on synthetic networks, each point is always an average over 100 different network realizations. We adopted NMI as a measure of consistency between the planted partition and the detected one.

\subsubsection*{Tests on the GN benchmark}
We first tested our method on the GN benchmark, which is the most famous benchmark for community detection. The GN network consists of 128 nodes which are divided into four equal groups. The edges are placed independently and randomly between node pairs, with probability $p_{in}$ for edges to fall between nodes in the same community, and $p_{out}$ for edges to fall between nodes in different communities. The values of $p_{in}$ and $p_{out}$ are chosen to make the expected degree of each node equal 16, and thus not independent. Here, we choose the mixing parameter $\mu$ as an independent parameter, which indicates the ratio of the external degree of a node with respect to its community to the total degree of the node.

\begin{figure}[ht]
	\centering
	\includegraphics[width=0.8\textwidth]{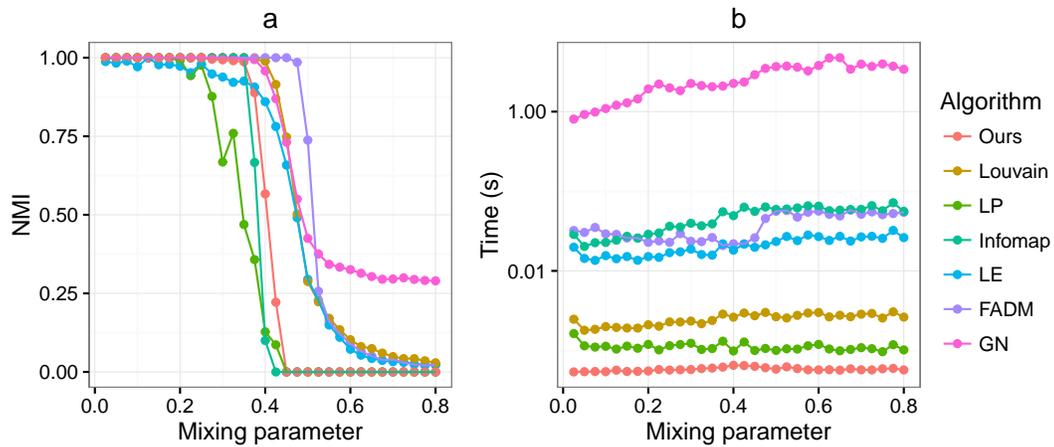}
	\caption{(Colour online) \textbf{Tests of the algorithms on the GN benchmark.} (a) shows the normalized mutual information as a function of the mixing parameter. (b) shows the execution time of different algorithms on the benchmark.}
	\label{GN} 
\end{figure}

Figure \ref{GN} (a) shows the average NMI scores of different algorithms. As can be seen, LP fails to detect the communities even for small $\mu$ ($\mu\sim 0.3$). Though better than LP, LE does not have a remarkable performance either, as it also starts to fail for low values of $\mu$. Our algorithm and Infomap have comparable performance. Both of them are better than LE, but outperformed by modularity-based methods: GN, FADM and the Louvain method. GN performs nearly as well as the Louvain method does. FADM performs the best, which yields a high NMI up until $\mu\sim 0.5$, where the rest algorithms cannot detect communities accurately.

Figure \ref{GN} (b) shows the execution time of different algorithms as a function of $\mu$. As we can see, our algorithm runs extremely fast, even faster than LP. This could be that the consideration of weights (or similarities) reduces the label oscillations compared with LP, which enables our algorithm to converge faster. The Louvain method is sightly slower than LP, but significantly faster than the rest of four methods. Infomap and FADM have comparable execution time, and both of them are sightly slower than LE. GN is the slowest among all algorithms.

\subsubsection*{Tests on the LFR benchmark}
We then tested our method on the LFR benchmark, and compared the results with those of the other methods. In the LFR network, both the degree of each node and the size of each community are drawn from power-law distributions. Each node shares on average a fraction $1-\mu$ of its edges with the other nodes of its community and a fraction $\mu$ with the nodes of the other communities; $0 \le \mu \le 1$ is the mixing parameter which indicates the significance of community structure.

\begin{figure}[ht]
	\centering
	\includegraphics[width=\textwidth]{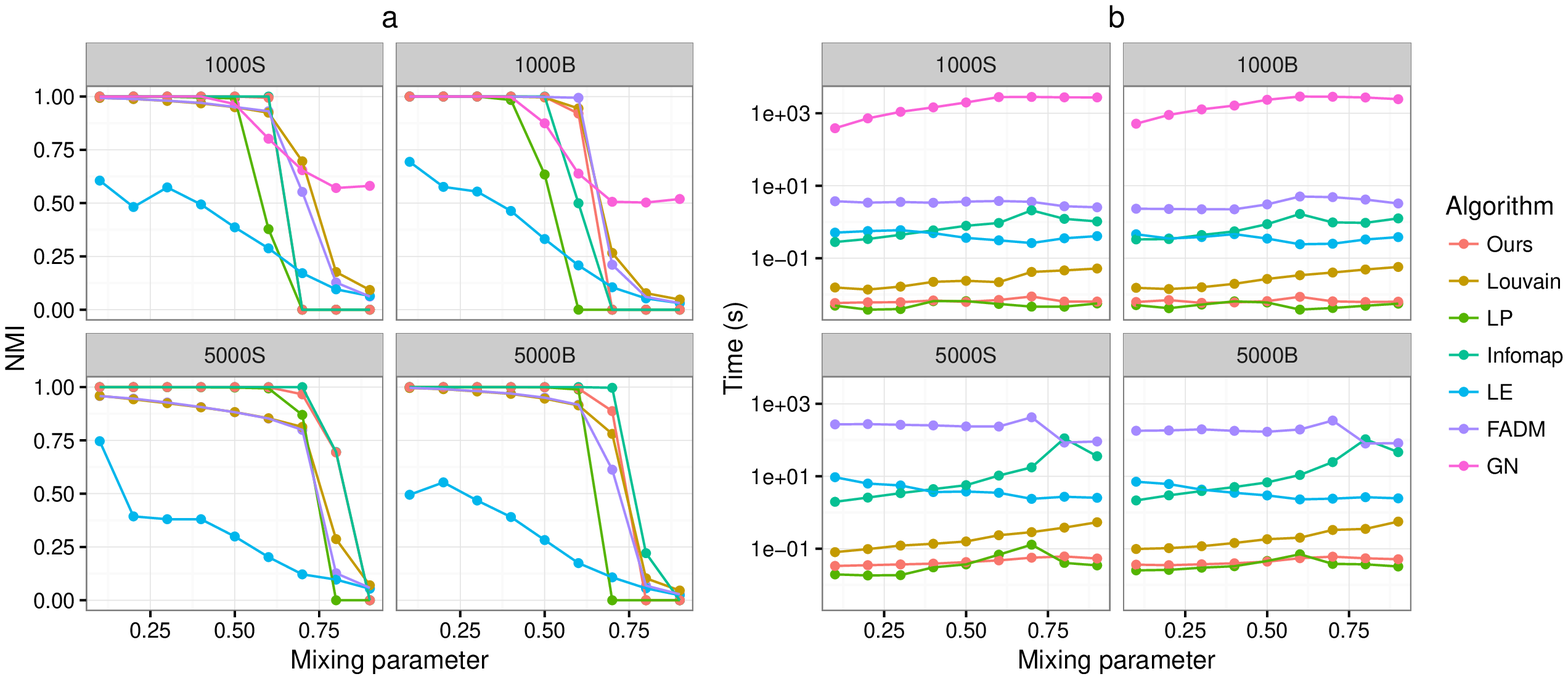}
	\caption{(Colour online) \textbf{Tests of the algorithms on the LFR benchmark.} (a) shows the normalized mutual information as a function of the mixing parameter. (b) shows the execution time of different algorithms on the benchmark.}
	\label{LFR} 
\end{figure}

In Figure \ref{LFR}, we show the average NMI and execution time of different algorithms as a function of the mixing parameter on the LFR networks. The following parameters apply to all LFR networks used here: the average degree and the maximum degree are 20 and 50 respectively, the exponents of the degree distribution and the community size distribution are -2 and -1 respectively. We show four sample results which correspond to two different network sizes (1000 and 5000), and two different ranges regarding community sizes, indicated by the letters `S' (small communities with 10 to 50 nodes) and `B' (big communities with 20 to 100 nodes), respectively. For the GN algorithm, we only show the results on smaller networks due to the high computational complexity of the method.

Generally, modularity-based methods, such as GN, LE, FADM and the Louvain method, have rather poor performance, especially in the case of larger networks with smaller communities, due to the well-known resolution limit of modularity \cite{Fortunato02012007}. LP does not have impressive performance either and its performance is sightly affected by the size of communities, i.e., it performs slightly better in the case of smaller communities than in the case of larger communities. Infomap performs sightly better than our algorithm on larger networks, and has comparable performance with our algorithm in the case of smaller networks and communities. However, it is outperformed by our method in the case of smaller networks with bigger communities. The results confirm that our algorithm has a reasonable performance on the LFR networks. Moreover, our algorithm runs extremely fast and is only sightly slower than LP (see Figure \ref{LFR} (b)).

\subsubsection*{Tests on random networks}
We also tested our method on random networks. In random networks, the connecting probabilities of the nodes are independent of each other, which leads to homogeneous density of links in the network. Thus, there should be no community structures. A good algorithm should not find non-trivial partitions and ideally deliver only one community which contains all nodes of the network.

Here, we considered two types of random networks: Erd\H{o}s-R\'{e}nyi\cite{erdds1959random} (ER) and scale-free\cite{Barabasi509} (SF). In ER random networks, nodes have the same probability to get connected to each other and the degree distribution is then binomial. The SF random networks are built via the configuration model \cite{molloy1995critical}, by starting from a certain degree sequence for the nodes obeying the predefined power law distribution with exponent -2. The sizes of all networks are fixed to 1000.

\begin{figure}[ht]
	\centering
	\includegraphics[width=0.8\textwidth]{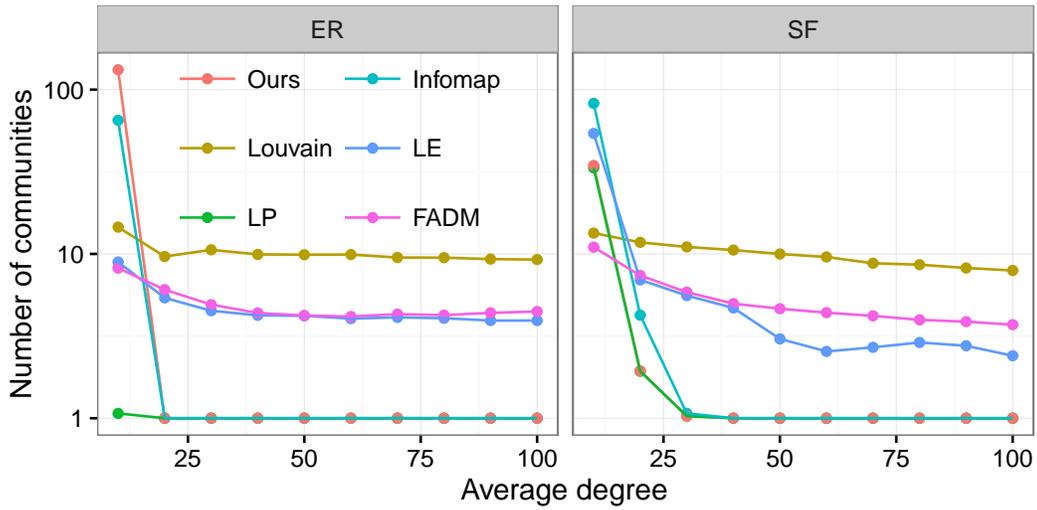}
	\caption{(Colour online) \textbf{Tests of the algorithms on ER and SF random networks.} The plots show the number of communities detected by different algorithms as a function of the average degree.}
	\label{ER_SF} 
\end{figure}

Figure \ref{ER_SF} shows the number of communities detected by different methods as a function of the average degree. The GN algorithm is excluded because it is too slow to be used for analysis. In both ER and SF random networks, LP, Infomap and our method always find a single community containing all nodes of the network except when the average degree is small. However, modularity-based methods, such as LE, FADM and the Louvain method, are not so good, as they always find a few communities even for a large average degree. These results show that our algorithm tends to find a few small communities in a sparse random network (due to stochastic fluctuations, specific realizations of random networks may display pseudo-communities), while it always detects a single community in a random network with dense connections.

\subsection*{Analysis of resolution limit}
\begin{figure}[ht]
	\centering
	\includegraphics[width=0.8\textwidth]{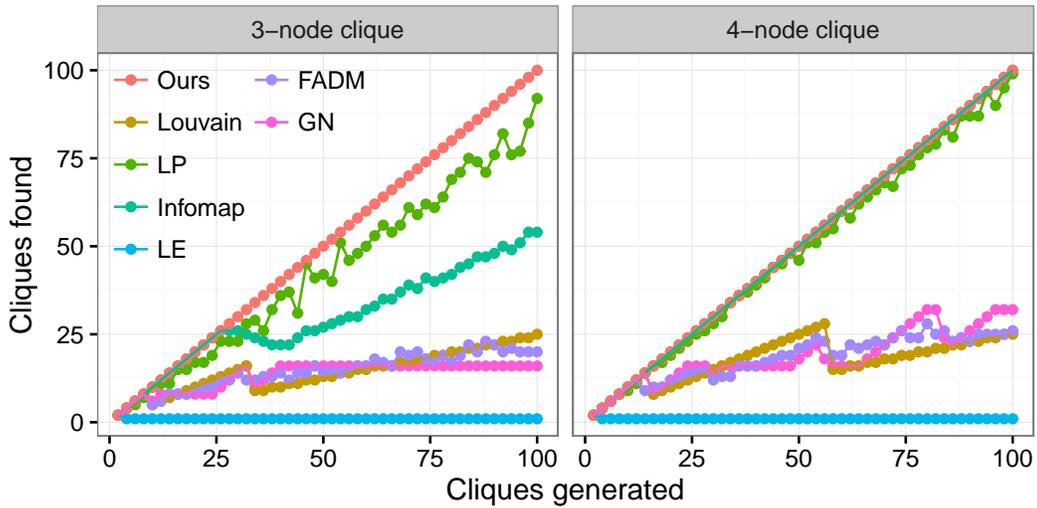}
	\caption{(Colour online) Tests of the algorithms on networks comprised of identical cliques which are connected by single edges.}
	\label{ResolutionLimitBench} 
\end{figure}

Finally, we analysed a kind of network made of some identical complete networks (cliques), connected by a single edge with each other, and each clique contains three nodes at least. Ideally, a good algorithm should detect all the predefined cliques. Figure \ref{ResolutionLimitBench} shows that our method detects all the cliques in all these examples. Infomap finds all 4-node cliques, but fails to detect all the 3-node cliques when the network gets large (i.e., with more than 26 cliques). LP finds sightly less communities than the predefined ones. The remaining four modularity-based methods detect dramatically less communities than the predefined ones when the network is large enough. This is mainly due to the resolution limit of modularity \cite{Fortunato02012007}, i.e., the optimization of modularity may fail to identify small communities below a cutoff size depending on the network size.

\subsection*{Real-World Networks}

The real-world networks are far more complex than the synthetic ones. Thus, it is still a great challenge to uncover the community structure of real-world networks. In this section, we applied our method to 9 different real-world networks listed in Table~\ref{realworldnetworks}. The sizes of these networks range from tens to millions. We presented the detailed results as follows.

\begin{figure}[ht]
	\centering
	\includegraphics[width=\textwidth]{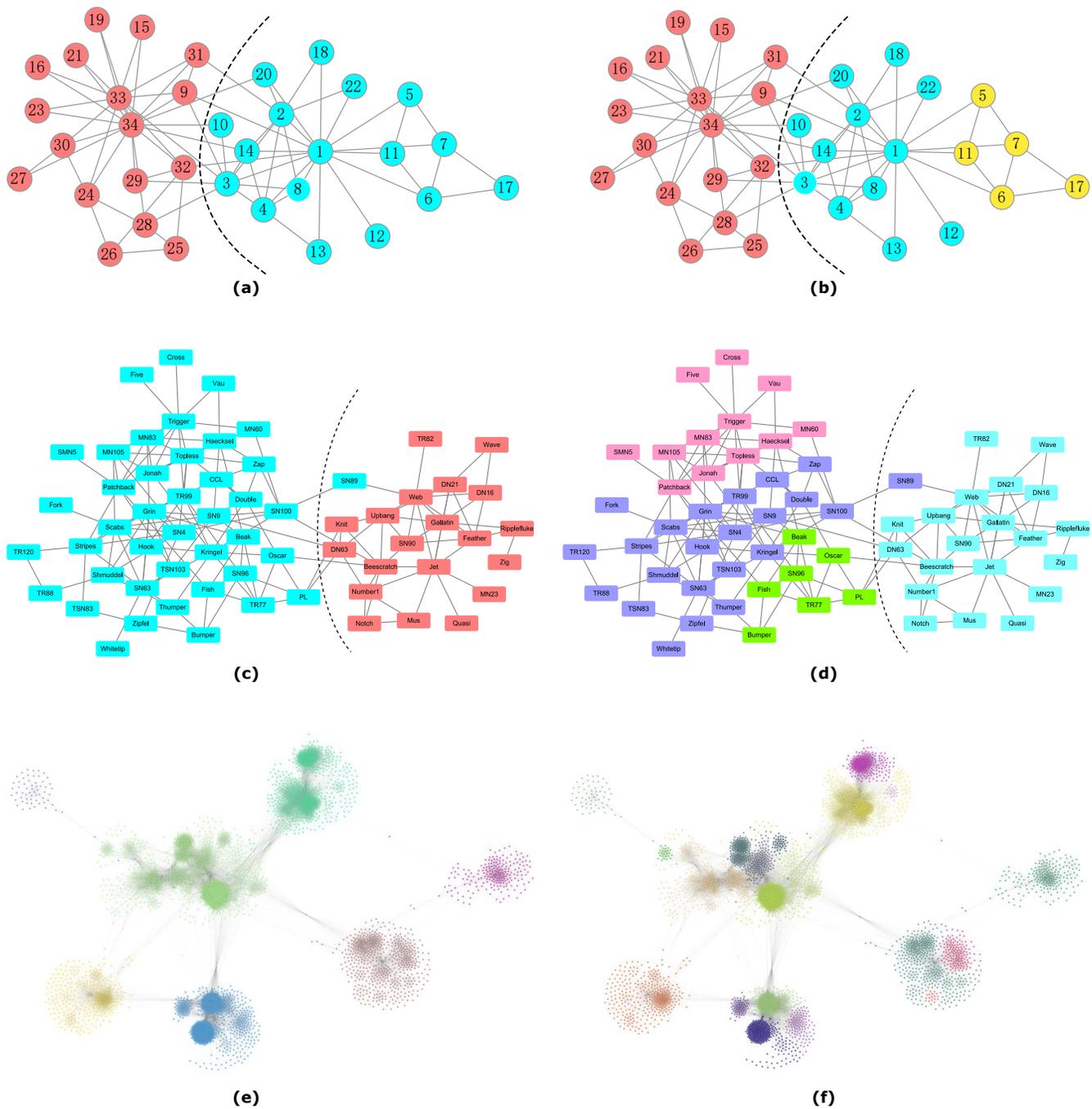}
	\caption{(Colour online) \textbf{Communities detected by our method on the real-world networks.} (Real communities are separated by the dashed curve and detected communities are distinguished by different colours). (a) Two communities discovered by our algorithm on the \emph{Karate} network with $\lambda=0.6$ are identical with the two real ones. (b) Three communities detected by our algorithm on the \emph{Karate} network with $\lambda=1$. (c) Two communities discovered by our algorithm on the \emph{Dolphins} network with $\lambda=0.6$, which are identical with the two real ones except for the node `SN89'. (d) Four communities discovered by our algorithm on the \emph{Dolphins} network with $\lambda=1$. (e) Seven communities obtained by our algorithm on the \emph{Facebook} network with $\lambda=0.05$. (f) Nineteen communities obtained by our algorithm on the \emph{Facebook} network with $\lambda=0.2$.}
	\label{realworldresult1} 
\end{figure}

Zachary's karate club is a social network of friendships between 34 members of a karate club at a US university in the 1970s. It was divided into two smaller clubs after a dispute between club president John (node 34) and instructor Mr. Hi (node 1). When $\lambda=0.6$, two communities are detected by our algorithm, which are identical to the two real ones, as shown in Figure~\ref{realworldresult1} (a). When $\lambda=1$, one of the two communities is divided into two smaller ones, as shown in Figure~\ref{realworldresult1} (b).

The dolphin social network describes the frequent associations between 62 dolphins living off Doubtful Sound, New Zealand. The links represent that dolphins are observed to stay together more often than expected by chance during the years from 1994 to 2001. Figure~\ref{realworldresult1} (c) shows the two communities identified by our algorithm with $\lambda=0.6$, which are identical to the two real ones except for the node `SN89'. When $\lambda$ increases to 1, one of the two communities is divided into three smaller ones, as shown in Figure~\ref{realworldresult1} (d).

\begin{figure}[ht]
	\centering
	\includegraphics[width=\textwidth]{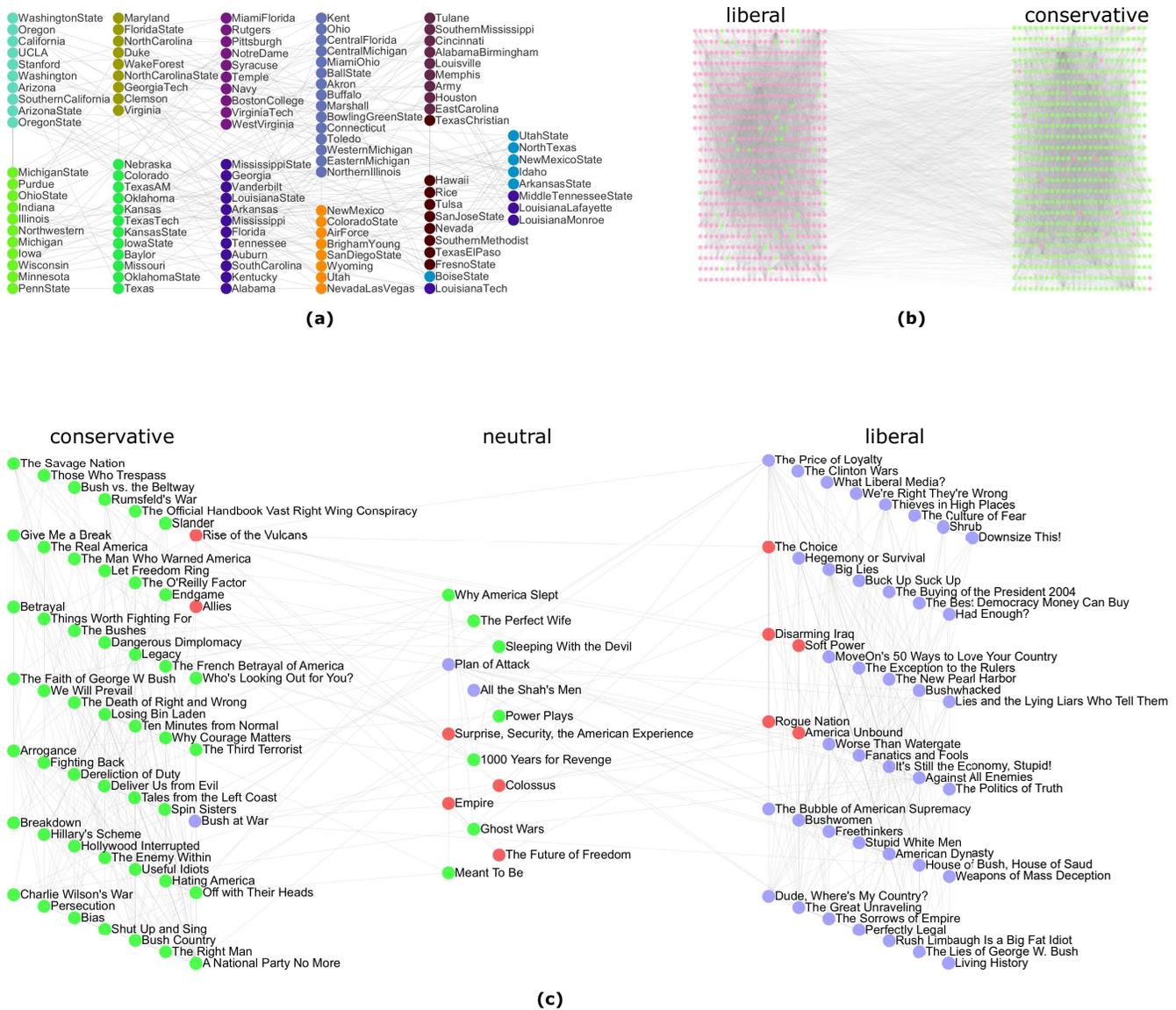}
	\caption{(Colour online) \textbf{Communities detected by our method on the real-world networks.} (The real communities are represented by the node positions and the detected communities are distinguished by different colours). (a) The \emph{Football} network and eleven communities obtained by our algorithm on it with $\lambda=0.6$. (b) The \emph{Polblogs} network: we discovered two large communities using our algorithm with $\lambda=0.6$. (c) The \emph{Polbooks} network and the three communities detected by our algorithm with $\lambda=0.6$.}
	\label{realworldresult2} 
\end{figure}

Books about US politics network is compiled by Valdis Krebs. Nodes represent books about US politics sold by the online bookseller Amazon.com. Edges represent frequent co-purchasing of books by the same buyers. Books have been divided into liberal, neutral, or conservative with respect to their attitudes. As shown in Fig.~\ref{realworldresult2} (c), our algorithm identifies three communities which resemble the three natural ones.

The American college football network describes football games among Division IA colleges during regular season Fall 2000. As shown in Fig.~\ref{realworldresult2} (a), 115 nodes in the network represent teams (identified by their college names), which are grouped into eleven different conferences, except for five independent teams (Utah State, Navy, Notre Dame, Connecticut and Central Florida). The regular season games between each pair of teams are shown as 613 edges of the network. When $\lambda=0.6$, our algorithm identifies elven communities within this network. Among them, eight conferences (i.e., Atlantic Coast, Big East, Big Ten, Big Twelve, Mid-American, Mountain West, Pacific Ten and Southeastern) are correctly identified. The three remaining communities closely resemble the Conference USA, Sun Belt and Western Athletic conferences. Five independent teams that do not belong to any conference tend to be grouped with the conferences which are most closely associated.

The E. Coli transcriptional regulatory network is a directed biological network with 578 edges and 423 nodes which is compiled by Shen-Orr et al. in 2002. Nodes are operons and edges start from an operon that encodes a transcription factor to another operon with which it directly regulates. Here we used an undirected version of the network described in the updated RegulonDB \cite{RegulonDB9}. We identified 5 isolated nodes, 26 modules with two operons, 9 modules with three operons and 22 modules with more than three operons (see Supplementary Fig.~S9 online). We analysed the 22 modules that have more than 3 elements with the DAVID functional annotation tool \cite{Huang2008,Huang01012009}. The greater probability that the genes appear to participate in a common biological process is described with smaller p-values. As shown in Table~\ref{annotations}, all these modules are functionally coherent. For example, the first module contains 53 operons, which performs as the cellular respiration (p-value is 5.2E-89). The second module has 6 operons and is involved in ``aromatic amino acid family biosynthesis process" (p-value is 3.2E-13). The results of other modules are listed in Table~\ref{annotations}. The entire operon list of the 22 modules can be found in Supplementary Table S2 online.

The \emph{Polblogs} network is a directed network of hyperlinks between weblogs on US politics, recorded in 2005 by Adamic and Glance. Here we use an undirected version of the network. As shown in Fig.~\ref{realworldresult2} (b), the resulting partition obtained by our algorithm mainly contains 2 communities, which indicates the liberal and conservative political leaning.

The \emph{Facebook} network describes friendships between users, which was collected from survey participants by using Facebook application. There are 4039 users and 88218 friendships in this network. We detected 7 communities by using our algorithm with $\lambda=0.05$ and 19 communities with $\lambda=0.2$, as shown in Fig.~\ref{realworldresult1} (e) and (f), respectively. The community structure of this network is quite clear based on visual observation.

\begin{figure}[ht]
	\centering
	\includegraphics[width=\textwidth]{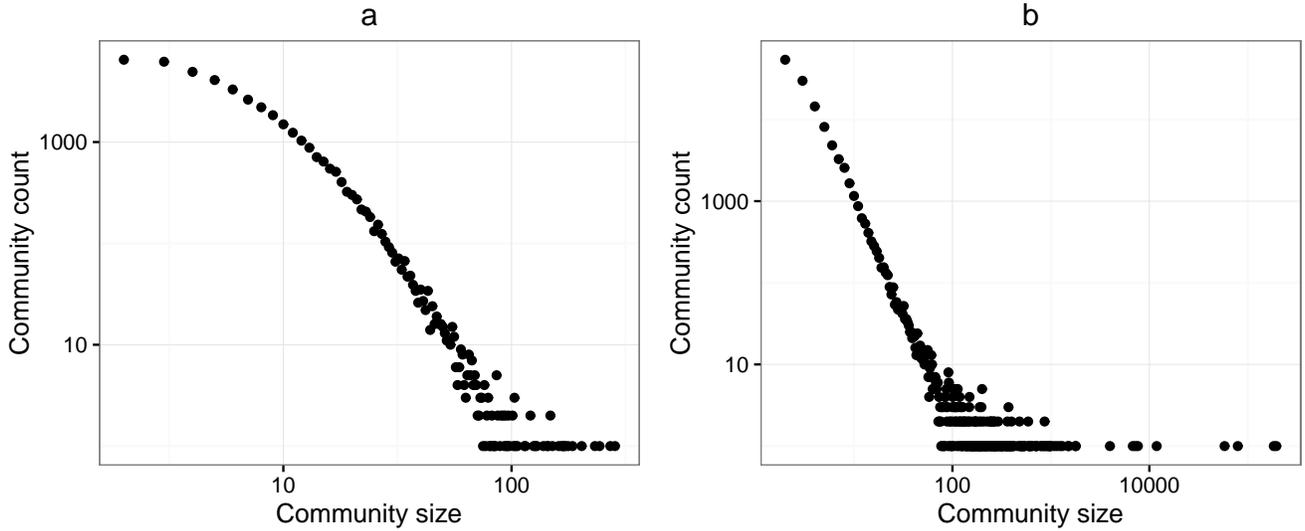}
	\caption{\textbf{Size distribution of detected communities.} (a) and (b) show the size distribution of communities detected by our algorithm with $\lambda=1$ on Amazon and Youtube networks respectively. In both cases, the behaviour is well reproduced by a power-law, and the exponents are -2.83 and -2.61 respectively.}
	\label{amazon_youtube_commdis} 
\end{figure}

The \emph{Amazon} network is collected by crawling the Amazon website. It is based on ``Customers Who Bought This Item Also Bought" feature of the Amazon website. If a product is frequently co-purchased with another product, the network contains an undirected edge between these two products. The Youtube social network contains the friendships of users. We analysed these two networks by using our algorithm with $\lambda=1$ and presented the distribution of community sizes in Figure~\ref{amazon_youtube_commdis}. It is found that the community sizes of both Amazon and Youtube networks can be well fitted by the power-law distributions, with exponents -2.83 ($x_{min}=10$, p-value=0.49) and -2.61 ($x_{min}=5$, p-value=0.45) respectively \cite{powerlaw}. This is consistent with the previous observations on other social networks \cite{PhysRevE.76.036106,PhysRevE.70.066111,PhysRevE.69.066133}.

\begin{figure}[ht]
	\centering
	\includegraphics[width=\textwidth]{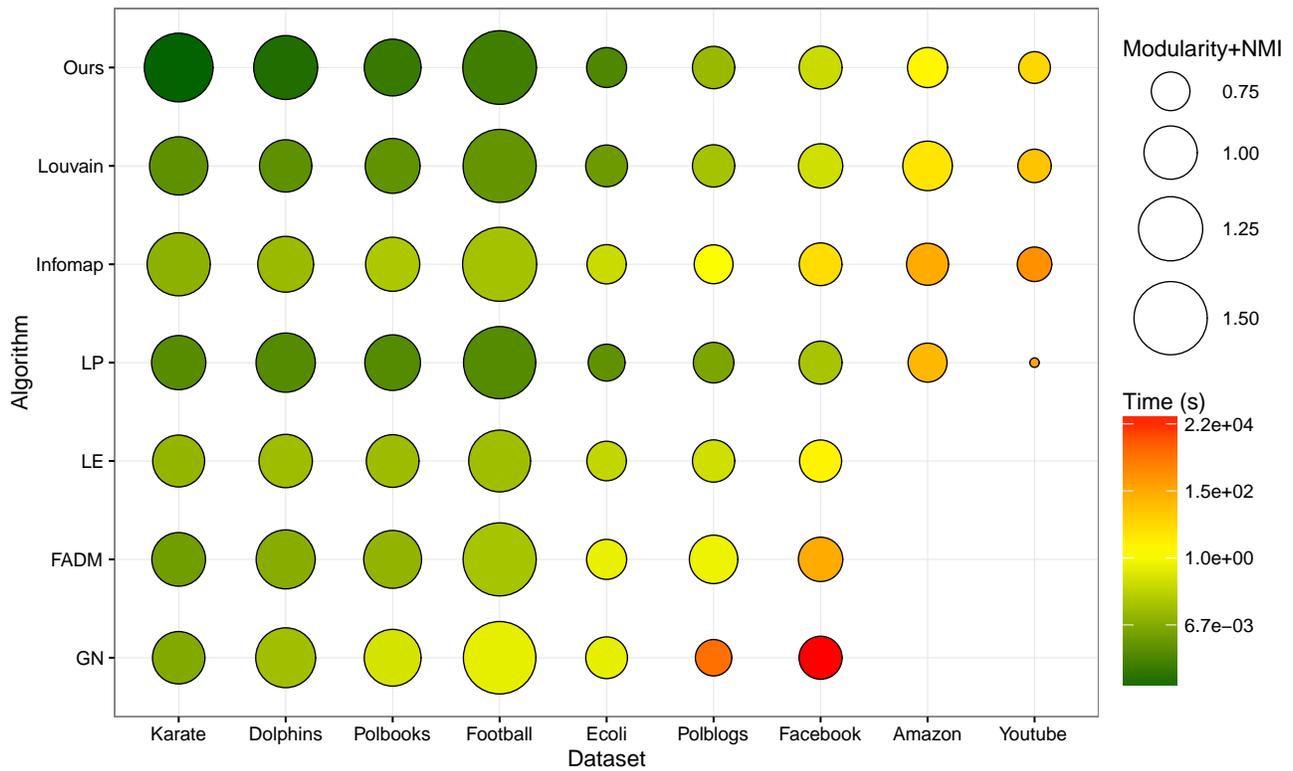}
	\caption{(Colour online) \textbf{Tests of the algorithms on real-world networks.} The bubbles show the composite performance and execution time for different algorithms. The composite performance includes two measures: NMI and modularity. For networks without ground truth partitions, we only show the modularity values. For LE, FADM and GN algorithms, the results are presented only for the smaller networks, due to the high complexity of the methods.}
	\label{RealWorldBench} 
\end{figure}

We also compared the results with those of the other methods, and showed their performances in Figure \ref{RealWorldBench}. The results of GN, LE, and FADM on some large networks are absent due to high computational cost. As can be seen, the composite scores (i.e., the sum of NMI and modularity) of our algorithm are competitive, especially on the \emph{Karate} and \emph{Dolphins} networks. Moreover, our algorithm runs extremely fast, even faster than LP on most networks. Therefore, our algorithm can detect communities on very large networks in short time with a reasonable accuracy.

\subsection*{Analysis of the resolution parameter}
\begin{figure}[ht]
	\centering
	\includegraphics[width=\textwidth]{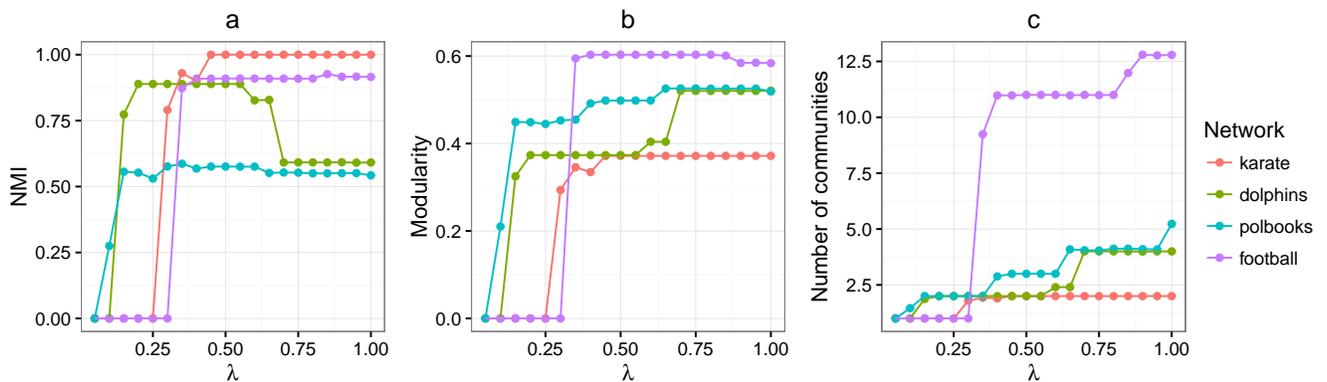}
	\caption{(Colour online) \textbf{Resolution parameter analysis on the four real-world networks.} (a), (b) and (c) show the NMI, the modularity, and the number of communities as a function of $\lambda$, respectively.}
	\label{RealWorldResolution} 
\end{figure}

In this subsection, we investigated the resolution parameter $\lambda$ in our method by studying four real-world networks, which are \emph{Karate}, \emph{Dolphins}, \emph{Polbooks} and \emph{Football} networks. As shown in Figure \ref{RealWorldResolution} (c), the number of detected communities will increase when $\lambda$ gets large. This is because the parameter $\lambda$ controls the weight of cohesion within a subnetwork. With the increase of $\lambda$, the importance of the cohesion will increase, and communities tend to be small and tight. While, when $\lambda$  is small, especially equal to 0, the attractions between subnetworks are always larger than their cohesions, thus the whole network will become one community eventually. In particular, the larger the resolution parameter, the smaller the detected communities.

In Figures \ref{RealWorldResolution} (a) and (b), we show the NMI and modularity of community structures at different scales. In general, the trend of community qualities for each network is like a stair, which increases from a low quality to a high value through some plateaus. This indicates the parameter $\lambda$ has some stable ranges, which could correspond to community structures of the network at different scales. In order to reveal the hierarchical structure of the network, we can employ a large resolution parameter firstly (e.g., $\lambda=1$), to obtain the community structure at low scales. Then, by decreasing the resolution parameter gradually, we can explore the community structures at high scales. Finally, the hierarchical structure of the network could be unfolded.

\section*{Discussion}
We propose a new method to detect community structures in complex networks. In our approach, similar nodes are first grouped together into meta-communities which will be retained or merged through a multilevel label propagation process (see \textbf{Methods} for details). We introduce a quantitative community criterion to determine what kind of subnetworks are communities. With the aid of this criterion, our algorithm can stop at an appropriate level when all the subnetworks meet our community criterion, and thus determine the number of communities automatically. Moreover, by tuning a resolution parameter, we can reveal the hierarchical organization of the network.

Compared with several other popular algorithms, our method has robust performance (in terms of NMI) on both synthetic and real-world networks with which ground truth is known, and the modularity scores are also competitive in most of the real-world networks. Moreover, our algorithm has a linear time complexity and runs extremely fast, which enables it to handle very large networks.

In our experiments, we observe that the modularity obtained by our method shows an increasing trend during the label diffusion process (see Supplementary Fig.~S3 online), although the maximization of modularity is not our intention. This provides further experimental evidence for the effectiveness of our method. In contrast to modularity-based methods, according to the testing results, our algorithm doesn't need to merge small communities to have a higher modularity, and thus preserves them even in large networks. Therefore, our method eliminates the resolution limit of modularity-based methods.

Moreover, our algorithm generally requires only a small number of iterations even on large networks (see Supplementary Fig.~S2 online). The number of meta-communities decreases dramatically in each iteration. Thus, most computing time is consumed in the calculation of similarities and the first iteration. In practice, if we want to explore the hierarchical structure of a network, we only need to calculate the similarities once, and then perform our algorithm on the network with different values of the resolution parameter. This makes our method faster in analysing the hierarchical organization of a network.

\section*{Methods}
\label{methods}
The key intuition behind our method is that, similar and tightly connected nodes are likely to belong to the same community. In the following subsections, we first define the similarity between two adjacent nodes, and introduce a criterion to determine what kind of subnetworks are communities. Then we present our algorithm in detail.

\subsection*{Structural similarity}
\label{similarity}
There are different measures based on the common neighbourhood to quantify the similarity of any pair of adjacent nodes. Here, we use the structural similarity measure given by the cosine similarity function \cite{Huang20112160}, which effectively denotes the local connectivity density of any two adjacent nodes in a weighted network. Given an undirected weighted network $G=(V,E,w)$, the structural similarity $s(u,v)$ between two adjacent nodes $u$ and $v$ is defined as:
\begin{equation}
s(u,v)=\frac{\sum_{x\in \Gamma(u)\cap\Gamma(v)}w(u,x)\cdot w(v,x)}{\sqrt{\sum_{x\in\Gamma(u)}w^2(u,x)}\cdot\sqrt{\sum_{x\in\Gamma(v)}w^2(v,x)}},
\label{similarity1}
\end{equation}
where $\Gamma(u)=\{v\in V|\{u,v\}\in E\}\cup \{u\}$ is the set containing $u$ and its adjacent nodes. If the network is unweighted, equation (\ref{similarity1}) can be simplified to be,
\begin{equation}
s(u,v)=\frac{|\Gamma(u)\cap \Gamma(v)|}{\sqrt{|\Gamma(u)|\cdot |\Gamma(v)|}}.
\end{equation}
$s(u,v)$ corresponds to the so-called edge-clustering coefficient introduced by Radicchi et al \cite{Radicchi2004}. The similarity value is in the range $(0,1]$. Two nodes will have a higher similarity by sharing more common neighbours. Once the neighbours of the two nodes are exactly the same, the similarity measure equals 1.

\subsection*{Community criterion}
\label{community}
We define a subnetwork as a community by comparing its internal and external connections (or weights). An undirected unweighted network $G$ can be represented by an adjacency matrix $A$ with entries $A_{uv}=1$ if $u$ is directly connected to $v$ and $A_{uv}=0$ otherwise. The subnetwork $V_i$ is a community if for any other subnetwork $V_j$
\begin{equation}
\lambda\sum_{u\in V_i, v\in V_i}A_{uv} \ge \sum_{u\in V_i,v\in V_j}A_{uv}.
\end{equation}
Informally, a subnetwork is a community if its internal connections exceeds the number of edges shared by the subnetwork with the other communities. Our definition of community is in the same spirit of weak community proposed by Hu et al \cite{PhysRevE.78.026121}. The difference is that we introduce a parameter $\lambda$ to quantitatively compare the internal and external connections (or weights) of each subnetwork. We can deem that $\lambda\sum_{u\in V_i, v\in V_i}A_{uv}$ is the cohesion of subnetwork $V_i$, and $\sum_{u\in V_i,v\in V_j}A_{uv}$ is the attraction between subnetworks $V_i$ and $V_j$. If $\lambda$ is very small, subnetworks tend to be attracted to each other and are merged to form large communities. On the contrary, when $\lambda$ is close to 1, subnetworks tend to form communities by their own. If $\lambda$ equals 1, the definition of weak community proposed by Hu et al. \cite{PhysRevE.78.026121} is recovered. Therefore, parameter $\lambda$ can be regarded as a resolution parameter which controls the scale upon which we would like to observe the communities in a network. Large $\lambda$ yields small, tight communities, and small $\lambda$ instead reveals large communities. In most situations, the whole network forms a single community when $\lambda$ approaches 0. In contrast, natural communities which are commonly detected by other algorithms are identified when $\lambda=1$.
	
\subsection*{Algorithm}
Our algorithm is based on the idea that communities are groups of nodes which are similar to each other, and communities can be detected from subnetworks by comparing the internal and external connections (or weights) of each subnetwork. The algorithm mainly consists of two steps: initialization and multilevel label propagation.

\textbf{Initialization}. It is observed that, the more similar two nodes are, the more likely they are assigned to the same community. So the similarity measure based on common neighbours provides a natural way to detect communities in networks. Therefore, in the initialization step, we first calculate the similarity between each pair of adjacent nodes. Then we re-assign labels to nodes in a way that each node takes the most similar label of its neighbours in a asynchronous manner. This label propagation process proceeds iteratively until the label of each node is one of the most similar labels in its neighbourhood. After the propagation process, we should obtain some meta-communities which consist of most similar nodes.

\textbf{Multilevel label propagation}. In general, the meta-communities detected in the initialization step do not all meet our community criterion, so we need the multilevel label propagation step to merge some of them. This step consists of two sub-steps: network collapse and label propagation, which are repeated iteratively.

Firstly, we built a new network whose nodes represent the meta-communities that were detected in the previous step. The weights of edges between the new nodes are given by the total weight of the edges between nodes inside the corresponding meta-communities. The total weight between nodes inside the same meta-community leads the weight of self-loop for this new node in the collapsed network. Once this sub-step is finished, we obtain a weighted collapsed network with weights representing the cohesion of meta-communities or attraction between meta-communities in the un-collapsed network.

Secondly, we perform a label propagation process on the collapsed network. To detect communities that meet the community criterion, we rescale the weights of self-loops by $\lambda$ prior to the label propagation process. Essentially, during the label propagation process, the meta-communities (i.e., the macro-nodes in the collapsed network) compare their internal cohesion with external attraction, and then decide to be retained or merged. After the label propagation, we should obtain some meta-communities of macro-nodes. Then it is possible to re-apply the first sub-step to collapse the network.

The number of meta-communities decreases dramatically in each iteration, and as a consequence most of the computing time is consumed in the first iteration. These two sub-steps are iterated until all the meta-communities meet our community criterion, i.e., the cohesion of each meta-community is greater than the attraction between it and any other meta-community. That is, meta-communities are no longer merged and thus the number of meta-communities no longer changes. Generally, the algorithm requires a small number of iterations (see Supplementary Fig.~S2 online). The detailed description of the algorithm and an illustration of the application of the algorithm to the \emph{Karate} network are shown in the supplementary material.
	
\subsection*{Evaluation measures}

To evaluate the effectiveness of the proposed algorithm we use the following two quality measures: modularity and normalized mutual information. As a widely used quality measure, modularity is defined by \cite{PhysRevE.69.026113}:
\begin{equation}
	\label{modularity}
	Q=\sum\left(e_{ii}-a_i^2\right),
\end{equation}
where $e_{ij}$ represents the fraction of total connections between two different communities, $e_{ii}$ represents the real fraction of links exist within a community, $a_i=\sum_j e_{ij}$ corresponds to the fraction of links connected to community $i$, and the expected number of intra-community links is just $a_i^2$. Modularity is based on the intuitive idea that random networks do not exhibit community structure. 
	
For a network with known community structure, consistency of the detected and the true partition is quantified by using the normalized mutual information \cite{1742-5468-2005-09-P09008}:
\begin{equation}
	\label{NMI2}
	I(A,B)=\frac{-2\sum_{i=1}^{c_A}\sum_{j=1}^{c_B}N_{ij}\log(N_{ij}N/N_{i.}N_{.j})}{\sum_{i=1}^{c_A}N_{i.}\log(N_{i.}/N)+\sum_{j=1}^{c_B}N_{.j}\log(N_{.j}/N)},
\end{equation}
where $A$ and $B$ represent the real and the detected partitions respectively. The number of real communities is denoted by $c_A$ and the number of detected communities is denoted by $c_B$. $N_{ij}$ is the number of nodes in the real community $i$ that appear in the detected community $j$. The sum over row $i$ of matrix $N_{ij}$ is denoted by $N_{i.}$ and the sum over column $j$ is denoted by $N_{.j}$. If the detected communities are identical to the real ones, $I(A,B)$ takes the maximum value 1. Whereas $I(A,B)=0$ if the partition detected by the algorithm is totally independent of the real partition.
	
\subsection*{Time complexity}
\label{complexity}
Given a network with $n$ nodes and $m$ edges. Let $k$ be the maximum degree of nodes in this network. The time complexity of each step of our algorithm is roughly estimated as follows.
\begin{enumerate}
	\item Similarity calculation takes time of $O(km)$ at most. For each pair of adjacent nodes, we need to iterate through all neighbours of the two nodes to calculate the similarity, and the upper bound of time complexity is $O(k)$. Thus calculating similarities for all pairs of adjacent nodes takes time of $O(km)$ at most.
	\item Aggregating network needs to iterate through each edge of the network, which takes time of $O(m)$ at most.
	\item Label propagation needs to iterate all nodes of the network, and each node iterates through at most $k$ neighbours, thus the upper bound of time complexity is $O(kn)$.
\end{enumerate}
Steps 2 and 3 repeat iteratively until all the communities meet the community criterion. Generally, our algorithm converges after a small number of iterations (see Supplementary Fig.~S2 online), so the time complexity is roughly $O(kn+km)$. If the network is sparse (i.e., $m\sim n$) and $k\ll n$, our algorithm has a linear time complexity, and thus can be efficiently applied to large-scale networks. 


\section*{Acknowledgements}

This work was in part supported by the Program of Introducing Talents of Discipline to Universities under grant no. B08033, and National Natural Science Foundation of China (Grant No. 11505071).

\section*{Author contributions statement}

J.H. designed the algorithm, carried out the simulations and prepared all the tables and figures. J.H., W.L. and W.D. prepared the main manuscript text. All authors reviewed the manuscript.

\section*{Additional information}

Competing financial interests: The authors declare no competing financial interests.

\begin{table}[ht]
	\centering
	\begin{tabular}{cccc}
		\hline
		Author and Ref. & Label & Time complexity \\ 
		\hline
		Blondel et al.\cite{FastUnfolding2008}  & Louvain & $O(m)$ \\ 
		Raghavan et al.\cite{PhysRevE.76.036106,igraph}   & LP (\textbf{L}abel \textbf{P}ropagation) & $O(m)$ \\ 
		Rosvall \& Bergstrom\cite{Rosvall29012008}   & Infomap & $O(m)$  \\ 
		Newman\cite{PhysRevE.74.036104,igraph}   & LE (\textbf{L}eading \textbf{E}igenvector) & $O(n^2)$  \\ 
		Treviño et al.\cite{1742-5468-2015-2-P02003}  & FADM (\textbf{F}ast and \textbf{A}ccurate \textbf{D}etermination of \textbf{M}odularity) & $O(n^3)$  \\
		Girvan \& Newman\cite{PhysRevE.69.026113,igraph}  & GN & $O(nm^2)$  \\
		\hline
	\end{tabular} 
	\caption{\label{algorithms} \textbf{List of the algorithms used in our experiments.} The first column indicates the designers and references, the second one denotes the label used to indicate the algorithm and the last one presents the time complexity of the algorithm. $n$ and $m$ indicate the number of nodes and the number of edges respectively.}
\end{table}

\begin{table}[ht]
	\centering
	\begin{tabular}{llll}
		\hline
		Dataset & Description & Nodes & Edges \\ 
		\hline
		Karate \cite{Zachary1977} & Zachary's karate club  & 34 & 78 \\ 
		Dolphins \cite{Lusseau2003} & Dolphin social network  & 62 & 159 \\ 
		Polbooks \cite{Krebs2008} & Books about US politics  & 105 & 441 \\ 
		Football \cite{NewmanSIAM2003} & American College football  & 115 & 613 \\ 
		E. Coli \cite{Shen-Orr2002} & Transcriptional regulation data & 423 & 519 \\
		Polblogs \cite{Adamic2005} & Weblogs on US politics (2005) & 1490 & 19090 \\
		Facebook \cite{NIPS2012_0272} & Facebook network & 4039 & 88218 \\
		Amazon \cite{Yang2013,snapnets} & Amazon co-purchased network & 334863 & 925872 \\
		Youtube \cite{Yang2013,snapnets} & Youtube friendship network & 1134890 & 2987624 \\
		\hline
	\end{tabular} 
	\caption{\label{realworldnetworks} A list of real-world networks employed in our experiments.}
\end{table}

\begin{table}[ht]
	\centering
	\begin{tabular}{llllll}
		\hline
		Index & $N_o$ & $N_g$ & Function clusters &  p-value & $N_m$\\ 
		\hline
		1&53&126&cellular respiration&5.20E-89&121\\
		2&6&7&aromatic amino acid family biosynthetic process&3.20E-13&7\\
		3&5&12&aromatic amino acid family biosynthetic process&2.00E-24&12\\
		4&8&23&sulfur metabolic process&1.60E-37&23\\
		5&8&17&iron transport&7.60E-46&17\\
		6&16&18&response to xenobiotic stimulus&7.50E-10&18\\
		7&11&11&stress response&2.10E-21&11\\
		8&15&23&purine biosynthesis&1.20E-31&23\\
		9&10&22&periplasm&1.00E-06&17\\
		10&7&13&signal&5.50E-10&13\\
		11&8&9&DNA binding&8.30E-11&9\\
		12&20&41&Lipopolysaccharide biosynthesis&2.40E-10&27\\
		13&5&6&fatty acid metabolic process&2.70E-09&6\\
		14&15&36&fermentation&1.30E-19&34\\
		15&13&53&flagellum&2.20E-97&41\\
		16&10&22&branched chain family amino acid biosynthetic process&1.50E-14&22\\
		17&7&7&methionine biosynthesis&1.50E-15&7\\
		18&12&16&SOS response&9.90E-28&16\\
		19&5&23&anion transport&5.10E-44&23\\
		20&26&55&oxidoreductase&6.80E-08&55\\
		21&72&144&carbohydrate catabolic process&1.30E-71&143\\
		22&7&10&arginine biosynthesis&8.10E-28&10\\
		\hline
	\end{tabular} 
	\caption{\label{annotations}\textbf{The function annotations of E. Coli. modules identified by our algorithm.} $N_o$ represents the number of operons in each module, $N_g$ represents the number of genes in each module, and $N_m$ represents the number of genes that are matched with the DAVID Escherichia coli database in each module.}
\end{table}

\end{document}